\documentclass[pra,twocolumn,10pt,aps,nofootinbib, superscriptaddress,longbibliography]{revtex4-2}  
\usepackage{graphicx}% Include figure files
\usepackage{dcolumn}% Align table columns on decimal point
\usepackage{bm}% bold math
\usepackage{amsmath}
\usepackage{mathtools}
\usepackage{graphicx}
\usepackage{color}
\usepackage{xcolor}
\usepackage{amsthm}
\usepackage{autobreak}
\usepackage{newlfont}
\usepackage{breakcites}
\usepackage{textcomp}
\usepackage{multirow}	
\usepackage{amssymb}
\usepackage{bm}
\usepackage{comment}
\usepackage[american]{babel}
\usepackage{braket}
\usepackage{soul}
\usepackage{float}
\usepackage{lipsum}  
\usepackage[normalem]{ulem} %to strike the words
\usepackage{hyperref}
\hypersetup{
     colorlinks  = true,
     linkcolor    = red,
     citecolor    = cyan,
     urlcolor     = magenta
     }
\begin{document}

\title{Strain topological metamaterials}
%FA*, RC*, DA*, IF, NB, FD, GT (*equal contributors)
\author{Florian Allein}
\thanks{These authors contributed equally}
%\email[]{florian.allein@univ-lille.fr}
\affiliation{Univ. Lille, CNRS, Centrale Lille, Univ. Polytechnique Hauts-de-France, Junia, UMR 8520-IEMN, F-59000 Lille, France}

\author{Adamantios Anastasiadis}
\thanks{These authors contributed equally}
\affiliation{Laboratoire d’Acoustique de l’Universit\'e du Mans (LAUM), UMR 6613, Institut d’Acoustique - Graduate School (IA-GS), CNRS, Le Mans Universit\'e, France}

\author{Rajesh Chaunsali}
\thanks{These authors contributed equally}
%\email[]{rchaunsali@iisc.ac.in}
\affiliation{Department of Aerospace Engineering, Indian Institute of Science, Bangalore 560012, India}

\author{Ian~Frankel}
\affiliation{Department of Mechanical and Aerospace Engineering, University of California, San Diego, La Jolla, CA 92093, USA}

\author{Nicholas Boechler}
\affiliation{Department of Mechanical and Aerospace Engineering, University of California, San Diego, La Jolla, CA 92093, USA}

\author{Fotios K. Diakonos}
\affiliation{Department of Physics, University of Athens, 15784 Athens, Greece}

\author{Georgios Theocharis}
\email[]{georgios.theocharis@univ-lemans.fr}
\affiliation{Laboratoire d’Acoustique de l’Universit\'e du Mans (LAUM), UMR 6613, Institut d’Acoustique - Graduate School (IA-GS), CNRS, Le Mans Universit\'e, France}

%%%%%%%%%%%%%%%%%%%%%%%%%%%%%%%%%%%%%%%%%%%%%%%%%%%%%%%%
%%%%%%%%%%%%%%%%%%%%%%%%%%%%%%%%%%%%%%%%%%%%%%%%%%%%%%%%

\begin{abstract}
Topological physics has revolutionised materials science, introducing topological insulators and superconductors with applications from smart materials to quantum computing. Bulk-boundary correspondence (BBC) is a core concept therein, where the non-trivial topology of a material's bulk predicts localized topological states at its boundaries. However, edge states also exist in systems where BBC is seemingly violated, leaving any topological origin unknown. For finite-frequency mechanical metamaterials, BBC has hitherto been described in terms of displacements, necessitating fixed boundaries to identify topologically protected edge modes. Herein, we introduce a new family of finite-frequency mechanical metamaterials whose topological properties emerge in strain coordinates for free boundaries. We show two examples, the first being the canonical mass-dimer, where BBC in strain coordinates reveals the previously unknown topological origin of its edge modes. Second, we introduce a new mechanical analog of the Majorana-supporting Kitaev chain. We theoretically and experimentally show that this Kitaev chain supports edge states for both free and fixed boundaries, wherein BBC is established in strains and displacements, respectively. Our findings suggest a previously undiscovered class of topological edge modes may exist, including within other settings such as electrical circuits and optics, and for more complex, tailored boundaries with coordinates other than strain.
\end{abstract}

\date{\today}
\maketitle

The field of condensed-matter physics has soared to new heights with the recent discovery of topological quantum matter. Topological insulating and superconducting materials, with the ability to support robust and defect-immune manipulation of electrons~\cite{Hasan2010, Qi2011, Bernevig2013, Sato2017}, have emerged as enabling candidates for the second quantum revolution~\cite{WenScience2019}. Numerous topological phenomena have also found their way from the quantum realm to the classical~\cite{Cooper2019, Ozawa2019, Susstrunk2016} despite the fundamental differences between electrons (fermions) and photons or phonons (bosons), and opened the way for new technologies relevant to optical, phononic, and mechanical computing \cite{Ozawa2019, Yasuda2021, Pirie2022}, and autonomous materials \cite{Pishvar2020}. 

An elegant combination of topological and band theory concepts relates the topological class of the bulk (infinite, periodic) system to the number of topologically robust, localized edge states on a finite sample's boundary. This connection between bulk topology and the existence of boundary states is commonly termed ``Bulk-Boundary Correspondence" (BBC)~\cite{helbig2020generalized,rhim2018unified,prodan2016bulk}. The BBC demands that the boundaries of a finite system do not break necessary symmetries for a topological classification \cite{ryu2010topological}. This framework naturally leads to a limited choice of boundary conditions for the emergence of topological boundary states. For example, in passive, finite-frequency topological mechanical metamaterials~\cite{huber2016topological, barlas2018topological, HuberNature2018} only fixed boundaries have been used to establish BBC~\cite{Wegener2019,2DGraphene,kariyado2016hannay}. Nevertheless, hidden symmetries can be revealed after suitable coordinate transformations \cite{andrianov1997matrix,li2015hidden,hou2013hidden,po2017symmetry,smith2019hidden,liu2022machine}. In this article we suggest that indeed topological systems with hidden chiral and particle-hole symmetries can exist, but the machinery of topology and BBC can be applied to them only after a suitable coordinate transformation and choice of boundary. The existence of these systems leads to important implications, not only because overlooked topological states may exist in current systems, but more importantly, because new topological systems with tailored boundaries may be designed.

\begin{figure*}[t!]
\includegraphics[width=\textwidth]{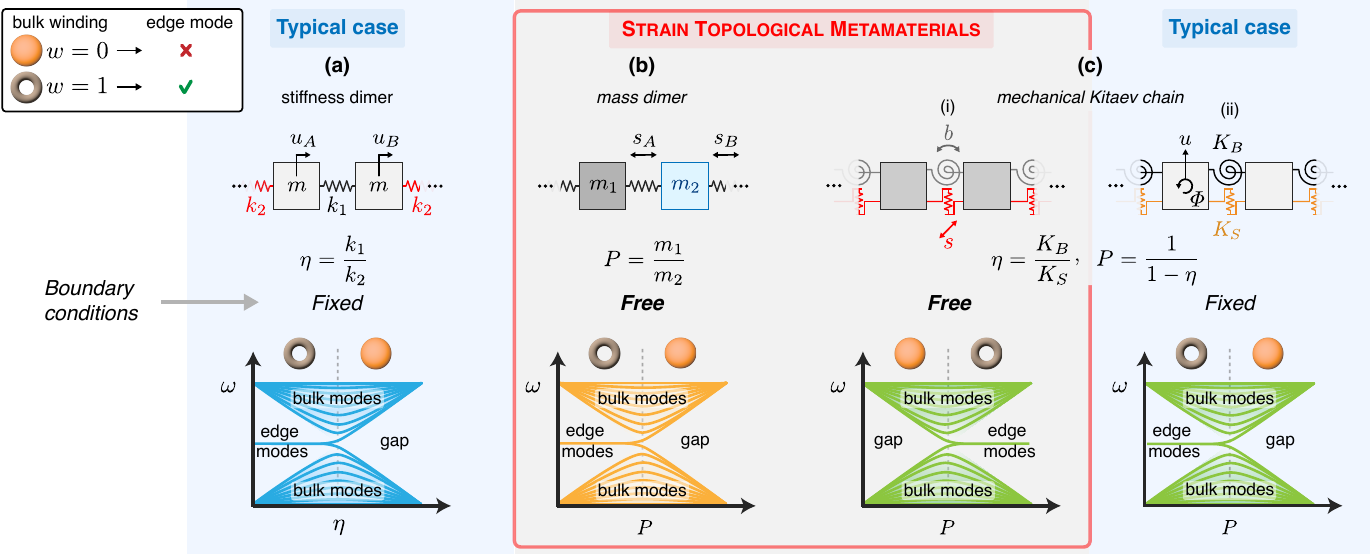}
\caption{
\textbf{Strain topological metamaterials.} A comparison of the new family of strain topological metamaterials (STM) to the typical case. In the top row, we present mass-spring schematics of different mechanical topological systems. In the middle row, we denote the appropriate boundary conditions for the existence of BBC. In the bottom row, we show how the spectrum of each system evolves while a parameter changes adiabatically. This is compared to the predictions of the bulk winding. The orange denotes trivial winding ($w=0$), and the doughnut non-trivial ($w=1$). \textbf{(a)} The stiffness dimer can be mapped to the finite-frequency SSH model and is a typical case. Only fixed boundaries preserve the chiral symmetry of the displacement bulk dynamical matrix $D_{u,\mathrm{bulk}}$. Edge states appear according to the prediction of the latter's winding number. \textbf{(b)} The mass dimer is an STM. As a result, its chiral symmetry is revealed only in strain coordinates, and edge states can exist only for free boundaries according to the winding number of the bulk dynamical matrix in strain coordinates $D_{s,\mathrm{bulk}}$. 
\textbf{(c)} The new mechanical Kitaev chain behaves both like an STM and a typical case, depending on the applied boundaries. (i) For free boundaries, it behaves like an STM, and the winding of $D_{s,\mathrm{bulk}}$ predicts the emergence of edge states correctly. (ii) For fixed boundaries, it behaves like a typical case, and the winding of $D_{u,\mathrm{bulk}}$ predicts the emergence of edge states correctly. Remarkably, the topological phases of this system are interchanged when different boundaries are applied.}
\label{fig1} 
\end{figure*}

Specifically, we show that mechanical finite frequency systems with free boundaries belong to a new family of topological mechanical metamaterials, whose topological properties become apparent not in the standard displacements coordinates ($\bm{u}$) but in strain coordinates ($\bm{s}$). Given a linear compatibility matrix $C$ defined as $\bm{s} = C\bm{u}$~\cite{kane2014topological}, we derive equations of motion in terms of bond extensions (strains) $\Ddot{\bm{s}} = - D_{s}\bm{s}$, where $D_{s}$ is the strain dynamical matrix. Using this, we extend the BDI class~\cite{ryu2010topological} of topological mechanical metamaterials to include systems whose bulk topology is probed by the winding number of the bulk $D_{s,\mathrm{bulk}}$  and show that BBC holds for free instead of fixed boundaries (the subscript ``bulk'' in $D_{s,\mathrm{bulk}}$ denotes the infinite - or equivalently periodic - system. If there is no such subscript we refer to the finite, bounded system with boundary conditions that break translation invariance). We call these systems \textit{strain topological metamaterials} (STM). For finite frequency metamaterials, the topological invariants can be defined for the shifted dynamical matrix~\cite{barlas2018topological,huber2016topological}, and in this work, we always imply -- if not else noted -- that the winding number is defined upon the shifted dynamical matrix.

For a complete topological characterization of a mechanical system, it is necessary to compare this new approach to the traditional one, where finite-frequency topological edge states are probed via the dynamical matrix $D_{u}$ defined in terms of lattice displacements where $\Ddot{\bm{u}} = - D_{u}\bm{u}$. For the set of fixed and free boundaries, three possibilities exist for finite-frequency topological mechanical systems, as outlined in Fig.~\ref{fig1}. The first possibility is systems that exhibit edge states only for fixed boundaries, where their topology is encoded in the winding number of the bulk $D_{u,\mathrm{bulk}}$. We will call this the typical case. An example of this is the mechanical Su–Schrieffer–Heeger (SSH) model \cite{Susstrunk2016,huber2016topological,PRL_Rajesh_SSH, TheocharisPRB2021} shown in Fig.~\ref{fig1} \textbf{(a)}. The second is systems that exhibit edge states only for free boundaries, where their topology is encoded in the winding number of $D_{s,\mathrm{bulk}}$. We use the mass dimer to demonstrate this case~\cite{Wallis1957, Allen2000}. The winding number of $D_{u,\mathrm{bulk}}$ is not well-defined in displacement coordinates, yet remarkably, $D_{s,\mathrm{bulk}}$ restores chiral symmetry, and the value of its winding corresponds to the emergence of edge states as we show in  Fig.~\ref{fig1}\textbf{(b)}. Finally, the third possibility is systems that exhibit edge states for both free and fixed boundaries, such that their topology is encoded in both $D_{u,\mathrm{bulk}}$ and $D_{s,\mathrm{bulk}}$, shown in Fig.~\ref{fig1}\textbf{(c)}. The model we use to demonstrate this is a mechanical analog of the Kitaev chain. The Kitaev chain has drawn particular interest for its support of Majorana modes, which have been suggested as promising candidates for quantum computing~\cite{kitaev2001unpaired,leumer2020exact}. We introduce a new mechanical analog, which facilitates the topological transition between trivial and nontrivial regimes. This system exhibits edge states for \textit{both} free (Fig. \ref{fig1}\textbf{(c)(i)}) and fixed boundaries (Fig. \ref{fig1}\textbf{(c)(ii)}), but for different values of its parameters. Our experimental results verify the ``double BBC'' predicted for this model, wherein we observe edge states for both fixed and free boundaries associated with the parameter values predicted by the winding of $D_{u,\mathrm{bulk}}$ and $D_{s,\mathrm{bulk}}$, respectively. 

While we examine only one-dimensional systems belonging to the BDI class, other classes of strain topological metamaterials may also exist, including nonreciprocal elements and in higher spatial dimensions. Apart from this, we emphasize that our findings are based on dynamical matrices of ``mass-spring" models, which are powerful tools to describe a plethora of other physical settings, including electrical circuits~\cite{Lee2018} and optics~\cite{palmer2022asymptotically}. We anticipate similar results therein. Further, localized topological states are known to exist at the interface between materials of two different topological phases, which raises the possibility of new classes of interface states. Lastly, our results suggest that similar methods can be applied to more complex, tailored boundary conditions with ``higher-order'' coordinates other than strain. 

\subsection*{Mass Dimer}
We begin with the mass dimer shown in Fig.~\ref{fig1}\textbf{(b)}. This system is a periodic 1D mass-spring chain with two alternating masses, $m_{1}$ and $m_{2}$, connected with a spring of stiffness $k$. The equations of motion of the particle displacements ($u_{A|B,n}$) in the $n_{th}$ unit cell are given by: 
\begin{align}
  m_{1}\Ddot{u}_{A,n} = k(u_{B,n}-u_{A,n}) - k(u_{A,n} - u_{B,n-1}
    )\label{eom1}\\
    m_{2}\Ddot{u}_{B,n} = k(u_{A,n+1}-u_{B,n}) - k(u_{B,n} - u_{A,n}
    ),\label{eom2}
\end{align}
where the first subscript denotes the sublattice within the unit cell and the second subscript $n$ denotes the unit cell number. We seek plane wave solutions of the form $\bm{\psi}_{n}(t) = \bm{u}(q) e^{i \Omega t - i q n}$, where $q$ is the normalized wavenumber and $\Omega$ the angular frequency. This results in the eigenvalue problem $D_{u,\mathrm{bulk}}(q) \bm{u}(q)  =  \omega^2 \bm{u}(q)$, where  $\bm{u}(q) = [u_A(q),  u_B(q)]^T$, $D_{u, \mathrm{bulk}}(q)$ is the Bloch dynamical matrix in displacement coordinates, and $\omega = \Omega/\Omega_0$ the normalized frequency with respect to the mid-gap frequency $\Omega^{2}_{0} = k(1/m_{1}+1/m_{2})$.
It is well known that the edge states of the mass dimer appear for free edges when the ratio $P:=m_1/m_2$ is varied~\cite{Allen2000}. However, their topological nature has been hitherto unknown, since the dynamical matrix $D_{u,\mathrm{bulk}}(q)$ lacks the necessary symmetries for a topological classification \cite{SI}.
 
\begin{figure}[t!]
\begin{center}
\includegraphics[width=1\columnwidth]{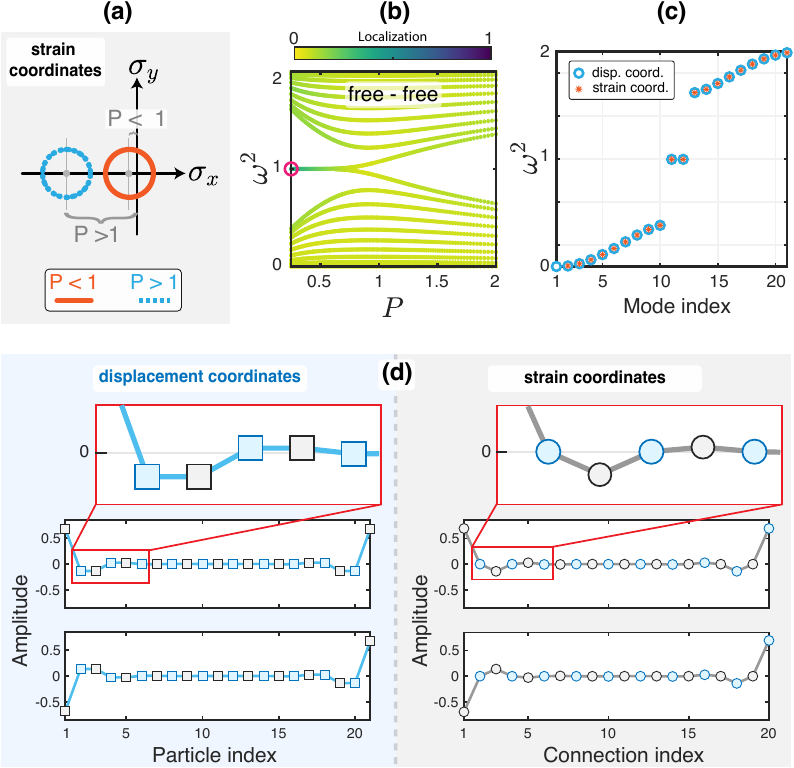}
\end{center}
\caption{\label{fig2} \textbf{Mass dimer.} {
\textbf{(a)} The chiral symmetry of mass dimer -- revealed on strain coordinates -- leads to a well-defined winding number. Nonzero winding makes the configurations with $P<1$ topologically nontrivial.
\textbf{(b)} Evolution of the spectrum of a finite chain with an odd number of particles (21) and free boundaries as we change the parameter $P$. Edge states emerge for $P<1$. Colormap confirms localization of states inside the band gap.
\textbf{(c)} Spectrum of a finite chain in displacement and strain coordinates at $P = 0.25$. Except for the zero mode, the spectrum is the same in both coordinates and shows chiral symmetry about the midgap frequency $\omega^2 = 1$.
\textbf{(d)} Profiles of the edge states in (c). Their chiral nature is revealed in strain coordinates.
}}
\end{figure}

We argue that the edge states in this model have a topological origin that can be revealed using strain coordinates. The strain coordinates for the $n$th unit cell are $s_{A,n} = u_{n,B} - u_{n,A}$ and $s_{B,n} = u_{n+1,A} - u_{n,B}$. Assuming plane wave solutions, we arrive at the following eigenvalue problem: $D_{s,\mathrm{bulk}}(q) \bm{s}(q)  =  \omega^2 \bm{s}(q)$, where  $\bm{s}(q) = [s_A(q),  s_B(q)]^T$, and $D_{s,\mathrm{bulk}}(q)$ is the Bloch dynamical matrix in \textit{strain} coordinates:  
\begin{align}\label{general_sm_st}
    D_{s,\mathrm{bulk}}(q) = \frac{1}{(1+P)}
    \begin{pmatrix} 
   1 + P & -(P+ e^{-iq})\\
    -(P + e^{iq}) & 1+P
    \end{pmatrix}.
\end{align}
The matrix $D_{s,\mathrm{bulk}}(q)$ can be written in terms of Pauli matrices $\sigma_x$, $\sigma_y$ and $\sigma_z$, such that ${D_{s,\mathrm{bulk}}}(q) = \bm{I} + d_x \sigma_x + d_y \sigma_y $ with $d_x =  (P+\cos{q})/(1+P)$ and $d_y =  \sin{q} /(1+P)$. As a result, the matrix anti-commutes with $\sigma_z$ after a constant shift in the diagonal: $\sigma_{z}(D_{s,\mathrm{bulk}}(q)-\bm{I})\sigma_{z}^{-1} = -(D_{s,\mathrm{bulk}}(q)-\bm{I})$. In other words, the shifted $D_{s,\mathrm{bulk}}(q)$ is chiral. Thus, the system has a well-defined winding number on the $\sigma_x-\sigma_y$ plane, as is shown in Fig.~\ref{fig2}\textbf{(a)}. The winding number predicts a topological phase transition at $P=1$, with $P>1$ and $P<1$ corresponding to trivial and non-trivial phases, respectively. Therefore, we expect BBC for the mass dimer -- but in strain coordinates.

Figure~\ref{fig2}\textbf{(b)} shows the spectrum of a finite chain with \textit{free} boundaries and an odd number of particles (which means an even number of bonds). We witness the emergence of edge states inside the band gap for $P<1$ (corresponding to the lighter mass on the boundaries), as expected by the strain winding number. We note that BBC dictates that the finite dynamical matrix, $D_{s}$, should also preserve the underlying chiral symmetry. 
This preservation is validated by the chiral operator for the finite matrix, which is defined as $\Gamma = \sigma_{z}\oplus\sigma_{z}\oplus...\oplus \sigma_{z}$ (see Methods).

%% FIGURE 3
\begin{figure*}[!]
\begin{center}
\includegraphics[width=1\textwidth]{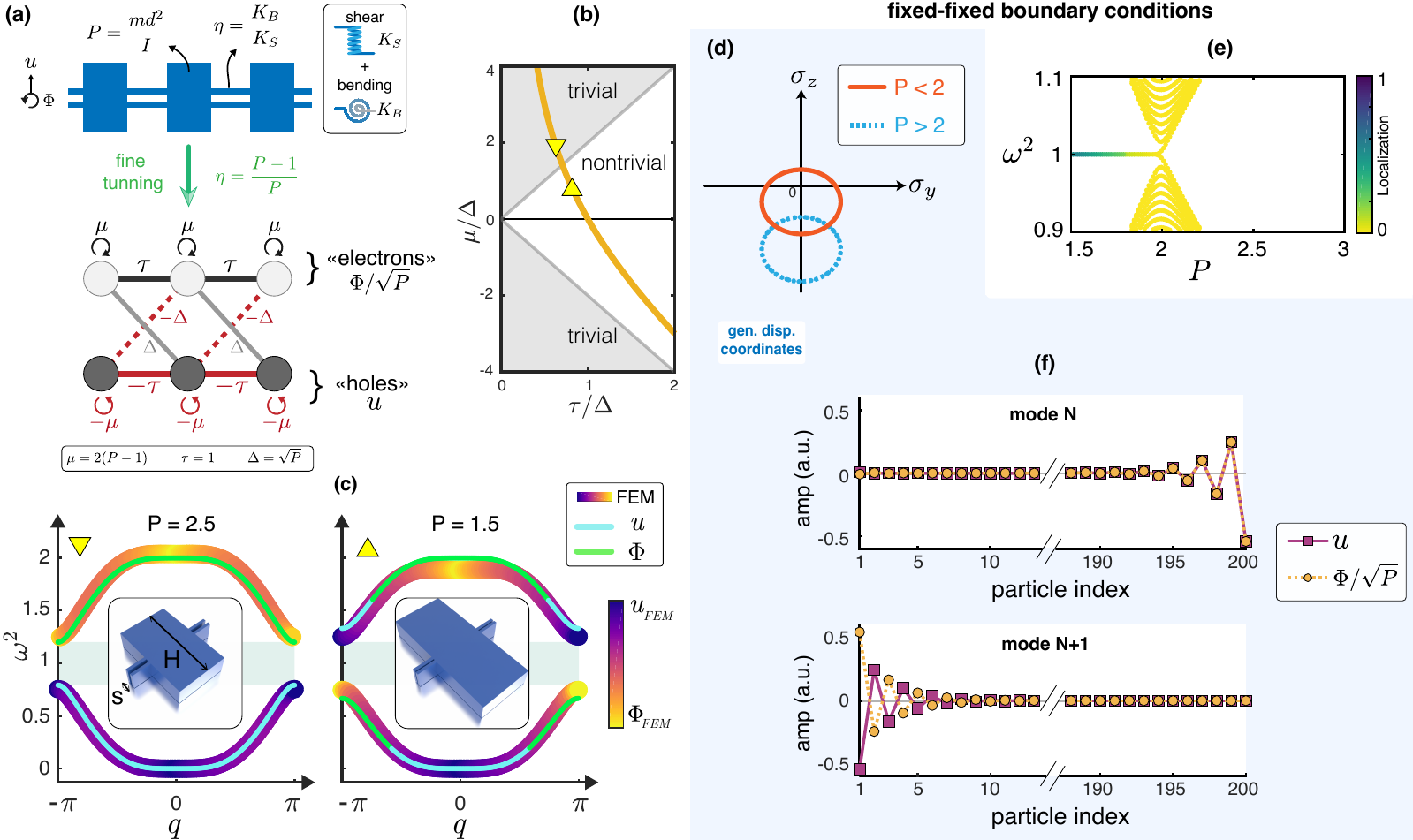}
\end{center}
\caption{\label{fig3} \textbf{Mechanical Kitaev chain.} 
 \textbf{(a)} A mechanical monomer chain with transverse and rotational degrees of freedom maps to the Kitaev chain after fine-tuning. 
\textbf{(b)} Topological phase diagram of the Kitaev chain. The path of the fine-tuned mechanical chain follows the curved solid yellow line. Two cases experimentally tested herein ($P=1.5$ and $P=2.5$) are marked with triangles.
\textbf{(c)} Dispersion diagrams for $P=1.5$ and $P=2.5$ are obtained in two ways: Analytically, via the lumped-mass model, and numerically, using the finite element method. $H$ and $s$ are the varying dimensions. The colorbar denotes modal dominance.
\textbf{(d)} The winding number of $\tilde{D}_{u,\mathrm{bulk}}$ suggests a topologically non-trivial phase for $P<2$.
\textbf{(e)} Evolution of the spectrum of finite chain with an even number of particles (200) and fixed boundaries  as we change $P$. Edge states emerge for $P < 2$. 
\textbf{(f)} Profiles of the edge states in (e) for $P=1.5$. Due to particle-hole symmetry, the profiles of effective particles $u$ and holes $\Phi/\sqrt{P}$ are either identical or differ by a phase.
}
\end{figure*}

We contrast these findings with the interpretation of the chain in the typical displacement coordinates. In  displacement coordinates, the chain has a zero-frequency mode, which corresponds to the rigid body motion of the free chain and breaks chirality. The strain description predicts all the nonzero eigenvalues of the system as is shown in Fig.~\ref{fig2}\textbf{(c)}. We witness the chiral symmetry of these non-zero eigenfrequencies with respect to the mid-gap frequency $\omega^2=1$. Furthermore, in Fig.~\ref{fig2}\textbf{(d)}, we show the profiles of the edge states at $P=0.25$ in both displacement and strain coordinates. Once again, strain coordinates reveal the chiral nature of the chain, where the vanishing amplitude of the topological edge states at alternating bonds is akin to the mechanical SSH (stiffness dimer)~\cite{Yang2021}. 

Building on the idea of BBC for strain coordinates and revealing new topological modes, we construct a mechanical analog of the Kitaev chain (the prototypical model for a topological superconductor) with two degrees of freedom per site. These degrees of freedom, specifically particle displacement and rotation, lead us to choose generalized strain coordinates involving both DOFs and probe the topological nature of the Kitaev chain. For the first time, we demonstrate that this design not only obeys BBC for fixed boundaries (right column of Fig.~\ref{fig1}\textbf{(c)}), but also shows a topological edge mode for free boundaries that can be explained by BBC in strain coordinates (left column of Fig.~\ref{fig1}\textbf{(c)}).

\subsection*{Mechanical Kitaev chain}
In Fig.~\ref{fig3}\textbf{(a)}, we show a mechanical structure whose dynamics are governed by two in-plane degrees of freedom (DOFs) at each site (transverse displacement $u$ and rotation $\Phi$). Each site is connected with the next via two bonds corresponding to bending and shear stiffness ($K_{B}$ and $K_{S}$, respectively). We set $P=md^{2}/I$, the ratio of the generalized masses (particle mass $m$, lattice constant $d$, and particle mass moment of inertia $I$) and $\eta = \frac{K_{B}}{K_{S}}$, the ratio of generalized stiffnesses (with $K_B$ and $K_S$ the bending and shear stiffnesses, respectively).

In Methods, we analytically show that if we impose the fine-tuning: $\eta = 1 - (1/P)$, the dynamical matrix on displacement coordinates $\tilde{D}_{u}$ maps to a Kitaev chain \cite{kitaev2001unpaired} (the ``$\sim$'' sign refers to the fine-tuned system). Parameter $P$ is mapped to the chemical potential $\mu$, the coupling $\tau$ and the superconducting constant $\Delta$ in the following manner: $\mu = 2(P-1)$, $\tau=1$, and $\Delta = \sqrt{P}$. As a result, transverse displacements $u$ and the normalized rotations $\Phi/\sqrt{P}$ can be seen as particle and hole DOFs (Fig. \ref{fig3}\textbf{(a)}). 
This mapping allows us to switch between trivial and non-trivial topological phases by continuously altering the value of $P$ while retaining the fine-tuning. Since we vary $P$ in our design, we trace a 1D path in the phase space of the Kitaev chain, as is shown in Fig.~\ref{fig3}\textbf{(b)}, wherein transitions between topologically trivial and nontrivial phases are possible.
In Fig.~\ref{fig3}\textbf{(c)}, we show the dispersion curves obtained for values of $P$ corresponding to systems in different topological phases. We observe two branches in the dispersion diagram as a result of the lumped-mass model having two DOFs, i.e., $u$ and $\Phi$, per mass. We also observe that the entire spectrum ($\omega^2$) is symmetric about a mid-axis, which is $\omega^2=1$. This is due to the particle-hole symmetry, such that 
$\sigma_x \left( \tilde{D}_{u,\mathrm{bulk}}(q) - \bm{I} \right) \sigma_x^T  = - \left( \tilde{D}_{u,\mathrm{bulk}}(q) - \bm{I} \right)$. Since $\tilde{D}_{u,\mathrm{bulk}}(q)$ maps to the Kitaev chain BdG Hamiltonian, a finite chain with boundaries that preserve the symmetry of the bulk (i.e., a chain with fixed boundaries) will exhibit topological edge states. 

In Fig.~\ref{fig3}\textbf{(d)}, we plot the winding of the Bloch vector of the shifted $\tilde{D}_{u,\mathrm{bulk}}(q)$ in the $\sigma_y$-$\sigma_z$~plane. This suggests the existence of edge states for $P<2$. Indeed, for a fixed chain consisting of 200 particles, two localized states emerge in the band gap for $P<2$ as one can see in Fig.~\ref{fig3}\textbf{(e)}. In Fig.~\ref{fig3}\textbf{(f)}, we plot these two eigenstates, which are localized on the left and the right end of the chain. The particle-hole symmetry of the model dictates that the particle and hole DOFs of the edge mode eigenstates either exactly match (symmetric) or have opposite phase (antisymmetric) \cite{leumer2020exact}. In contrast to the edge states appearing in the SSH model~\cite{TheocharisPRB2021}, these topologically-protected edge states have mixed polarization in terms of displacement and rotation.

%% FIGURE 4
\begin{figure}[t!]
\begin{center}
\includegraphics[width=1\columnwidth]{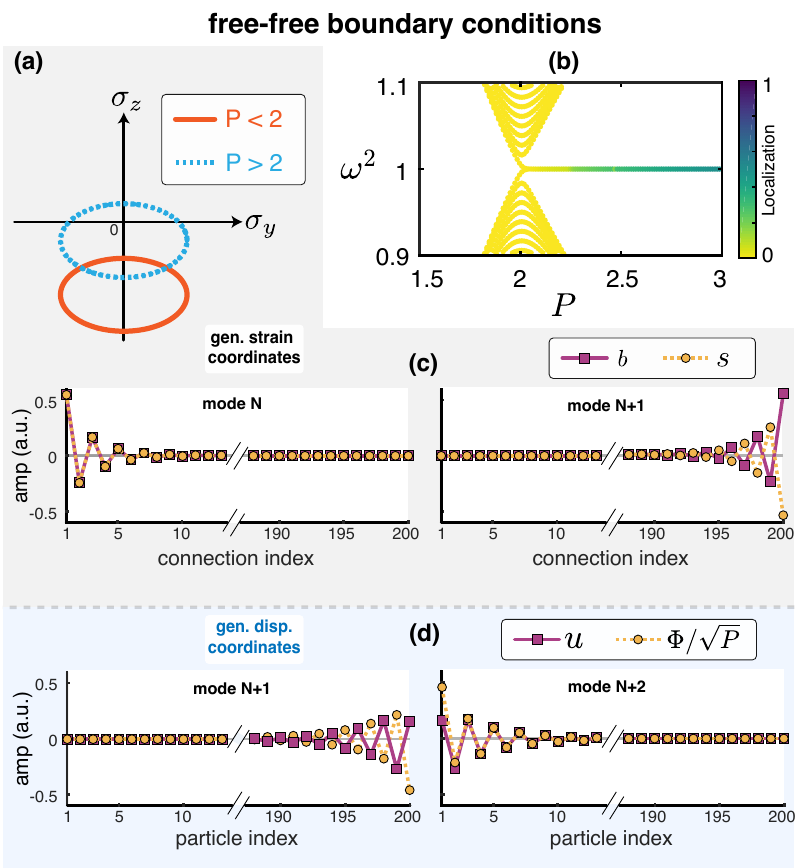}
\end{center}
\caption{\label{fig4}
\textbf{Analysis of the Kitaev chain with free boundaries. (a)} The winding number of $\tilde{D}_{s,\mathrm{bulk}}$ predicts a non trivial phase for $P>2$ contrary to the winding of $\tilde{D}_{u,\mathrm{bulk}}$ which predicted a non trivial phase for $P<2$ (see Fig. \ref{fig3}\textbf{(d)} for comparison). \textbf{(b)} Spectrum of a finite chain ($N=200$) with both boundaries free as a function of $P$. Localized edge states emerge in the band gap for $P>2$. 
\textbf{(c)} Profiles of the localized states on the free edges on strain coordinates. The bending ($b$) and shear ($s$) strain coordinates follow the pattern dictated by particle-hole symmetry. 
\textbf{(d)} Profiles of the localized states on the free edges on displacement coordinates. Their profiles appear distorted. 
}
\end{figure}

We now investigate the Kitaev system in strain coordinates, which now have a more complex form due to the coupling of rotational degrees of freedom with the transverse displacements. By applying the strain coordinate transformation, we obtain the bulk strain dynamical matrix $\tilde{D}_{s,\mathrm{bulk}}(q)$, which surprisingly maps \emph{again} to a Kitaev chain (as long as the fine-tuning is preserved) but with a different parameter dependence. In strain coordinates, $P$ is now replaced by $P/(P-1)$ (see Methods). In Fig.~\ref{fig4}\textbf{(a)}, we show the winding of $\tilde{D}_{s,\mathrm{bulk}}(q)$ predicts the inverse topological phases from those predicted by $\tilde{D}_{u,\mathrm{bulk}}(q)$. While a finite chain with fixed boundaries preserves particle-hole symmetry in displacement coordinates, we need a finite chain with \textit{free} boundaries in order to preserve particle-hole in strain coordinates~\cite{SI}.
As a result, we expect the emergence of edge states for a Kitaev chain with free boundaries but for the opposite parameter regimes compared to the system with fixed boundaries (topologically nontrivial regimes become trivial and \textit{vice versa}).  

In Fig.~\ref{fig4}\textbf{(b)}, we show the spectrum of the Kitaev chain with free boundaries with varied $P$. Remarkably, we witness the emergence of two edge states inside the band gap for $P>2$, as predicted by the winding of the strain dynamical matrix $\tilde{D}_{s,\mathrm{bulk}}(q)$. These hidden topological edge states exhibit the profile dictated by particle-hole symmetry when expressed in the strain coordinate system (Fig. \ref{fig4}\textbf{(c)}), while their form appears distorted when expressed in displacement coordinates (Fig. \ref{fig4}\textbf{(d)}). Building off our unique definition of generalized strain in the Kitaev chain may also open the door for establishing symmetries based on other coordinates paired with the appropriate boundaries.

\subsection*{Experimental results}

To experimentally verify our predictions, we prepare a test setup to probe the Kitaev system with fixed-free boundary conditions so that both types of edge states can be observed in the system without changing the mounting. For a large chain with a negligible interaction between two boundaries, we expect the emergence of an edge state at the fixed end, as dictated by the BBC of the fixed-fixed chain. Similarly, we expect an edge state at the free end as well, albeit for different $P$ values than the fixed chain.
As such, the fixed-free chain should always have an edge state at one edge for all values of $P$ except $P=2$ (where the band gap closes). For systems with $P<2$ and $P>2$, they would support an edge state on the fixed and free ends, respectively.

%% FIGURE EXPERIMENTAL RESULTS
\begin{figure}[t]
\begin{center}
\includegraphics[width=\columnwidth]{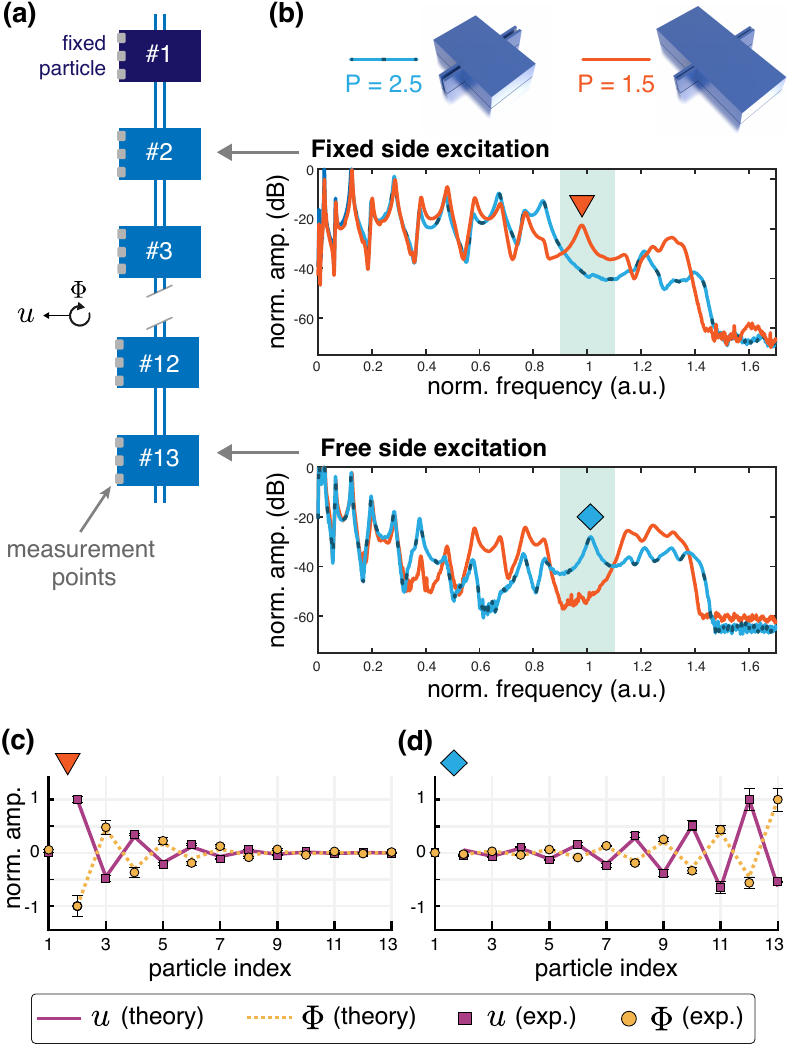}
\end{center}
\caption{\label{fig5} 
\textbf{Experimental observation of edge states in the Kitaev chain.}
(a) Schematic of the experimental setup, suspended vertically by fixing particle $\sharp$1.
Three points are probed on each particle to characterize the transverse displacement and rotation.
(b) Measured frequency response at particle $\sharp$7 when the chains with $P=1.5$ and $P=2.5$ are excited at the fixed end (at particle $\sharp$2) or at the free end (particle $\sharp$13). The blue area corresponds to the band gap. 
Measured amplitudes of the edge state (displayed in displacement coordinates) (c) localized at the fixed boundary for $P=1.5$, and (d) at the free boundary for $P=2.5$.}
\end{figure}

We fabricate chains of 13 masses (large cuboids) through additive manufacturing and suspend them vertically by mounting the particle $\sharp1$ as shown in Fig.~\ref{fig5}\textbf{(a)}. Therefore, the system represents a fixed-free chain. We consider two chains with different $P$ and excite them with an automatic modal hammer by striking the particle $\sharp2$ or $\sharp13$ corresponding to the fixed or free sides. By using a laser Doppler vibrometer, we then measure the velocity at multiple points along the chain. See Methods and Supplementary Information~\cite{SI} for more details on fabrication, experimental setup, and data acquisition.

Figure~\ref{fig5}\textbf{(b)} shows the experimentally measured frequency response at particle $\sharp7$ when the chains with $P=2.5$ and $P=1.5$ are excited from different ends. We witness a band gap (highlighted region) and a peak inside it, which appears for a given chain and side of excitation, corresponding to the edge state. The state inside the band gap exists at the fixed end for $P=1.5$ and at the free end for $P=2.5$, as theoretically predicted.

To verify that these modes are indeed localized at different edges, we reconstruct the mode shapes from the experimental data in Figs.~\ref{fig5}\textbf{(c,d)}. We observe excellent agreement between predictions and experiments, where amplitude decay can be seen as one goes away from the boundaries. We also note that the edge state localized at the free end [Fig.~\ref{fig5}\textbf{(d)}] is different in its shape compared to its counterpart for the fixed edge, as discussed earlier, corroborating the inversion of topological phases predicted for our mechanical system. 

\def\bibsection{\section*{}} 
\bibliography{bibliography.bib}

\section*{Acknowledgements}
We thank N. Herard, B. Skoropys, and M. Coimbra for earlier investigation on the experimental realization of a mass-spring system via additive manufacturing.
R.C. acknowledges the funding support by the Science and Engineering Research Board (SERB), India, through the Start-up Research Grant SRG/2022/001662.
I.F. acknowledges support from the NDSEG Fellowship Program through the US Army Research Office.
N.B. acknowledges support from the US Army Research Office (Grant No. W911NF-20-2-0182).
A.A. acknowledges support from the Greek State Scholarships Foundation (I.K.Y) as part of Nikolaos D.~Xrysovergis grant.

\section*{Author contributions}
F.A.,  A.A., and R.C. contributed equally to this work.
A.A. conceived the strain-description and developed the theoretical framework under F.D. and G.T.'s supervision. 
R.C. analyzed the fine-tuning for the mechanical Kitaev chain and developed the theory on displacement coordinates.
F.A., I.F., and N.B. conceived the general beam modeling. F.A. designed the experimental samples, performed the experiments, and analyzed the data.
All authors contributed to the conceptualization of the project and the writing of the manuscript.

\section*{Competing interests}
The authors declare no competing interests.

\end {document}